\documentstyle[12pt]{article}
\newcommand{\beq}{\begin{equation}}
\newcommand{\eeq}{\end{equation}}

\def\half{{\textstyle{1\over2}}}
\def\quart{{\textstyle{1\over4}}}
\def\kap{{\textstyle{1\over{\kappa^2}}}}

\def\p1half{{\textstyle{{{p+1}\over{2}}}}}

\def\23phalf{{\textstyle{{{23-p}\over{2}}}}}

\begin{document}
\thispagestyle{empty}
\begin{titlepage}

\bigskip
\hskip 3.7in{\vbox{\baselineskip12pt
}}

\bigskip\bigskip\bigskip\bigskip
\centerline{\large\bf
Bosonic Matrix Theory and Matrix Dbranes}

\bigskip\bigskip
\bigskip\bigskip
\centerline{\bf Shyamoli Chaudhuri\footnote{On Unpaid Leave of Absence from
Penn State, Jan--Jul 2002.  Current E-mail: shyamolic@yahoo.com}}

\medskip
\medskip
\medskip

\centerline{214 North Allegheny Street}
\centerline{Bellefonte, PA 16823, USA}

\bigskip

\date{\today}

\bigskip\bigskip
\begin{abstract}
\noindent 
We develop new tools for an in-depth study of our recent proposal for Matrix Theory. 
We construct the anomaly-free and finite planar continuum limit of the ground state 
with $SO(2^{13})$ symmetry matching with the tadpole and tachyon 
free IR stable high temperature ground state of the open and closed bosonic string. 
The correspondence between large $N$ limits and spacetime effective actions is 
demonstrated more generally for an arbitrary D25brane ground 
state which might include brane-antibrane pairs or NS-branes and which need not have an 
action formulation. Closure of the finite $N$ matrix Lorentz algebra nevertheless 
requires that such a ground state is simultaneously charged 
under all even rank antisymmetric matrix potentials. Additional invariance under 
the gauge symmetry mediated by the one-form matrix potential requires a ground 
state charged under the full spectrum of antisymmetric ($p$$+$$1$)-form matrix 
potentials with $p$ taking any integer value less than 26. Matrix Dbrane democracy 
has a beautiful large $N$ remnant in the form of mixed Chern-Simons couplings in 
the effective Lagrangian whenever the one-form gauge symmetry is nonabelian. 
\end{abstract}
\noindent

\end{titlepage}

\section{Introduction}

\vskip 0.1in
The pre-eminent task facing string theorists of our time is finding the
answer to the deep, and puzzling, question: \lq\lq What is String Theory?" 
\cite{whatis}. We need an answer that is plausible, consistent with all 
of the known facts about weak-strong-dual effective field theory limits of 
nonperturbative String/M theory, and that is both mathematically, and 
aesthetically, satisfactory 
\cite{mthy,stfldth,bfss,dvv,ikkt,malda,kth}. In recent 
work, we have proposed a new matrix framework for nonperturbative string/M 
theory \cite{spont}. In what follows, we explain the motivation underlying
our proposal, illustrating many points within the context of the simpler 
bosonic matrix theory. In particular, we also include in this paper a 
worldsheet analysis of the open and closed unoriented bosonic string 
which is to be matched with the $1/N$ expansion about a planar 
limit of bosonic matrix theory. We will develop new tools necessary 
for an in-depth study of the proposed matrix framework for M theory
given in \cite{spont}. 

\vskip 0.1in
For the most part, our discussion focuses in this paper on {\em Bosonic Matrix 
Theory}.\footnote{We should clarify at the outset that our proposal has no 
relation to the intriguing conjecture put forth in \cite{bosm}.}
The inequivalent multiple-scaled large $N$ states of the quantum matrix 
effective action are conjectured to be in one-to-one correspondence with the 
nonperturbative states of the quantum open and closed bosonic string. Recall that 
unoriented, open and closed, bosonic string theory is the unique bosonic string 
that is both tadpole-free, and with nonabelian gauge fields in the critical 
spacetime dimension \cite{polbook,poltorus,anomopen}. There is a tachyonic 
mode in both the $SO(2^{13})$ and $SO(2^{12})$$\times$$SO(2^{12})$ bosonic string 
ground states at zero temperature. The tachyon can, however, be removed at 
temperatures above a critical temperature of order the string scale \cite{deconf}: 
$\beta$$<$$\beta_c$, where 
$\beta^2_c$$=$$4\pi^4\alpha^{\prime}/(\Delta w)^2$, in the presence of a time-like 
Wilson line. $T_c$ is the self-dual temperature of the bosonic closed
string theory. The $\Delta w$$=$$\pm 1$ states may be interpreted in the $T$-dual
coordinate as winding modes in Euclidean time. The stretched strings 
are massive at temperatures below the string scale, where the string vacuum has a 
tachyonic instability--- a runaway direction in the tachyon potential. 

\vskip 0.1in 
At $T$$=$$T_c$, however, the stretched strings turn massless enhancing the gauge 
symmetry to $SO(2^{13})$. Above $T_c$ the scalar field has positive mass, and 
$T$$=$$T_c$ is therefore a {\em metastable} turning point of the tachyon potential. 
This is the prediction of classical string theory. It is this high temperature 
stable ground state of bosonic string theory, both tadpole and tachyon free, which
will be the model of interest for us in this paper. We will conjecture that the 
quantum string vacuum with $SO(2^{13})$ enhanced gauge symmetry is the stable IR 
limit of the classically metastable open and closed bosonic string, at both zero 
and at finite temperatures. Some motivation for this identification will
be given, based both on prior work, and on expectations from open string field 
theory \cite{stfldth}. We perform matching calculations between the anomaly-free 
and IR stable high temperature ground state of the bosonic string and the planar 
limit of the quantum matrix effective action, in a ground state with $SO(2^{13})$
symmetry arising from $2^{12}$ matrix D24branes. We emphasize that the matrix theory 
calculations will give an unambiguous prediction for the higher derivative interactions 
in the effective action of the quantum open and closed bosonic string.\footnote{We 
note that if, instead, the metastable behavior of the classical bosonic string 
ground state is also found in the planar continuum limit of the bosonic matrix 
action, this would provide a nice example of a model with an inflationary potential. 
Modulo the issue of finding analogous phenomena in a realistic supersymmetric 
matrix ground state with Standard Model symmetries, and at suitable energy scales,
this is potentially a phenomenon of great interest \cite{deconf,tachyft}.} 

\vskip 0.1in
The worldsheet analysis of the classically metastable open and closed bosonic 
string at one-loop order is introduced in section 2 of this paper, moving on
to the high temperature stable quantum ground state of the bosonic string theory 
with $SO(2^{13})$
Yang-Mills gauge symmetry. Section 3 begins 
with a discussion of classical matrix actions, and the implementation of the 
Noether procedure for the bosonic finite $N$ matrix algebras. In particular, we 
discuss the construction of manifestly covariant matrix actions for 
$U(N)$$\times$$SL(d,R)$$\times$$\cal G$, where $\cal G$ is the group of extended 
Yang-Mills symmetry at finite $N$, and $d$$=$$26$, $10$, respectively, in the
bosonic, or super, matrix model cases. $\cal G$ includes symmetries mediated by the 
higher rank antisymmetric $p$-form matrix potentials. These symmetries are
described in some detail in section 3.2, moving on to a discussion of the
matrix quantum effective action for an arbitrary Dbrane state in section 3.2.

\vskip 0.1in
Matrix Dbrane states demonstrate a remarkably simple and elegant phenomenon
we will refer to as {\em Dbrane democracy:} closure of the 
finite $N$ matrix Lorentz algebra on any matrix theory ground state with D25brane 
charge {\em requires} that the ground state is simultaneously charged under all 
even rank antisymmetric matrix potentials.\footnote{This is distinct, although
not unrelated, to the use of the term {\em $p$-brane democracy} 
in \cite{town}.}
Additional invariance under the gauge 
symmetry mediated by the one-form matrix potential requires a ground state charged 
under the full spectrum of antisymmetric ($p$$+$$1$)-form matrix potentials with 
$p$ taking any integer value less than 26. Matrix Dbrane democracy has a beautiful 
large $N$ remnant in the form of mixed Chern-Simons couplings in the effective 
Lagrangian whenever the one-form gauge symmetry is nonabelian. This is described in
section 3.3. In the conclusions, we comment on the match between higher derivative 
terms in the $1/N$ expansion of the quantum matrix effective action and the 
$\alpha^{\prime}$ expansion of the quantum string effective action, clarifying also 
the physics intuition underlying large N reduced models \cite{ek,ikkt,spont}. We explore 
future directions of interest in relation to our work. 

\vskip 0.1in
In closing, we would like to apologize in advance for the absence in this paper of 
motivation and review of prior work, especially in the area of matrix theory
techniques. This 
is due to circumstances which prevented the author from following developments in this 
field on a regular basis. We hope that this odd absence in the introduction is adequately 
made up for, in part, by our referencing of papers we have found stimulating, and of 
possible relevance to our work. We offer our sincere apologies to those authors whose 
work has been thus overlooked. 

\section{Quantum Open and Closed Bosonic String Theory}

We begin with a review of the unoriented open and closed 
bosonic string theory \cite{polbook}. The one-loop vacuum amplitude 
for the tadpole-free unoriented string with gauge group $SO(2^{13})$ is given
in detail in Chap.\ 7 of Polchinski's text \cite{polbook}. In the modern 
language of Dbranes, this vacuum can be interpreted as that with $2^{12}$
coincident D25branes, whose worldvolume is the Minkowskian signature
spacetime with $(1,25)$ noncompact dimensions. In the tadpole-free vacuum,
the one-loop vacuum amplitude may be written in the simple
form \cite{polbook}:
\begin{equation}
{\cal A} = i V_{26} (2\pi^2 \alpha^{\prime})^{-13}
\int_0^{\infty} {{dt}\over{4t}} t^{-13}\eta(it)^{-24} \left ( 1 + 2 \mp 
  \eta(it)^{12} \Theta_{00}(0;it)^{-12} \right ) \quad .
\label{eq:amp1}
\end{equation}
For clarity, we have separated the contributions from Klein bottle (1)
and annulus (2) inside the round brackets. It is obvious from the $q$
expansion of this expression, where large $t$ gives the lowest-lying
states in the open string sector, that the spectrum will be tadpole free.
But the leading contribution from Klein bottle and annulus contains a
tachyonic state, contributing $3e^{2\pi t}$ to the $q$ expansion. This
is the result for the, classically unstable, $26$-dimensional open and 
closed bosonic string at zero temperature.\footnote{Since the classical
bosonic string at zero temperature is anyway unstable, one may ask why 
we are focused on
a tadpole-free background. The reason is that we are interested in 
finding a mechanism with clear analog for the supersymmetric type
I string. There, the dilaton tadpole {\em must} vanish in order to 
remove an accompanying unphysical tadpole for a Ramond-Ramond eleven
form (see the discussion in \cite{anomopen,polbook}). Moreover, as
explained below, the classical bosonic string is actually
{\em metastable} at finite temperature.}
 
\vskip 0.1in
Consider compactifying one dimension on a circle of radius $R$. For a 
spatial compact dimension,
we can include a spacelike Wilson line thus breaking the nonabelian gauge
symmetry to $SO(2^{12})$$\times$$SO(2^{12})$. The $T_{25}$-dual picture 
is a configuration of spatially separated stacks of $2^{11}$ D24branes, 
lying on orientifold planes at $X^{\prime}_{24}$$=$$0$, and 
$X^{\prime}_{24}$$=$$\pi R^{\prime}$. The $T$-dual radius is given by
$R^{\prime}$$=$$\alpha^{\prime}/R$. Consider the background with a stretched 
Dstring linking the well-separated D24brane stacks. The stretched strings
contribute a term of classical origin in the formula for the
open string mass
spectrum, and the mass of the tachyonic mode will be shifted for spatial 
separations much larger than the string scale. The one-loop amplitude in
this background is given by:
\begin{equation}
{\cal B} = i V_{25} (2\pi^2 \alpha^{\prime})^{-25/2}
\int_0^{\infty} {{dt}\over{4t}} t^{-25/2} \eta(it)^{-24} \left 
 ( 1 + 2 \mp \eta(it)^{12} \Theta_{00}(0;it)^{-12} \right ) 
 q^{ R^{\prime 2} /\pi^2 \alpha^{\prime} }
\quad ,
\label{eq:amp2s}
\end{equation}
and where it is understood that the spatial radius is much larger than
string scale distances. If $R^{\prime}$ is reduced to string scale 
values, the stretched strings turn massless, contributing additional
gauge bosons that enhance the symmetry to the full $SO(2^{13})$. For 
all separations above $R^{\prime}$$=$$R_c$, the tachyon is massive. The 
divergence due to the tachyon is recovered below $R_c$, there is a 
runaway potential for the tachyonic state. We emphasize that these 
are the predictions from the classical string theory analysis we are 
able to perform at zero temperature. Our interest here is in results
that hold at weak, not necessarily vanishing, open and closed string 
couplings. 

\vskip 0.1in
Consider the finite temperature behavior of this theory in the limit of 
well-separated D24branes. A Wick rotation to Euclidean signature 
with compact \lq\lq time" yields 
the finite temperature effective action functional, ${\cal W}(\beta)$,
for open and closed bosonic string theory at one-loop order for this 
string state. The reader is referred to the discussion given in 
\cite{deconf}, based on the earlier works cited in that paper. Besides 
the zero temperature modes described above, there are Matsubara thermal
modes, all of which are tachyonic at low temperatures, turning massless 
at given critical temperature of order the string scale. 
The free energy, $F(\beta)$$\equiv$$-{{1}\over{\beta}} W(\beta)$, at 
one-loop order in this background is given (in part) by an expression 
$F_1$ of the form \cite{deconf}:
\begin{equation}
F_1 = \beta^{-1} \int_0^{\infty} {{dt}\over{4t}} 
(2\pi t)^{-12} \eta(it)^{-24} \left 
 ( 1 + 2 \mp {{\eta(it)^{12} }\over{ \Theta_{00}(0;it)^{12}}} \right ) 
\sum_{n \in {\rm Z} , \Delta w = 0, 1} q^{ R^{\prime 2} /\pi^2 \alpha^{\prime} 
+ \alpha^{\prime} \pi^2 n^2 /\beta^2 }
\quad ,
\label{eq:amp2tt}
\end{equation}
where it is understood that the D24brane stacks are well-separated, and 
the spatial separation, $R^{\prime}$, is much larger than string scale.
Upon lowering the separation $R^{\prime}$ to $R^{\prime}_c$, the 
stretched strings will turn massless, contributing additional gauge bosons 
which enhance the gauge symmetry to $SO(2^{13})$. At low temperatures, the
tachyon will re-appear in the limit of coincident Dbranes. This would seem to
be a bit contrived, since we are accustomed to thinking of the Dbrane 
separation, or the compactification radius of the T-dual theory,
as a freely varying modulus. In particular, the simple-minded prescription 
for the finite temperature string state given here leads to the unphysical, 
and inescapable, conclusion that there are low temperature tachyons even in the
case of supersymmetric string theories. Fortunately, the finite temperature 
prescription has since been refined in our recent work \cite{deconf}, 
and we will use these papers for guidance in the discussion below. 

\vskip 0.2in
Notice that the expression in Eq.\ (\ref{eq:amp2tt}) is clearly incomplete, 
since it does not 
explained what happens to the finite temperature behavior of the $SO(2^{13})$ 
theory with coincident D24branes, the analog of the $SO(32)$ type I$^{\prime}$ 
string at finite temperature. As explained by us in \cite{deconf}, the finite 
temperature theory in either case, i.e., with well-separated, or coincident, 
D24brane stacks, is well-defined upon incorporating a timelike Wilson line, 
$\beta A^0$$=$$(1^8 , 0^8)$. This step simply corresponds to quantization of the
Yang-Mills theory in a modified axial gauge. The timelike component 
of the Yang-Mills potential, $A_0$,
has been set to a temperature-dependent constant, rather than zero. It 
implies that the nonabelian gauge group at finite temperature is 
$SO(2^{12})$$\times$$SO(2^{12})$. The circle radius, $R_{24}$, or its $T$-dual 
D24brane separation, $R^{\prime}_{24}$, is now a free modulus, as in the zero
temperature string, precisely as one would expect on physical grounds. 

\vskip 0.2in
In the presence of the time-like Wilson line, there are timelike winding modes 
stretched between the well-separated D24brane stacks, as in the analogous type
I$^{\prime}$ example described in \cite{deconf}. Let us denote the stacks by 
$w$$=$$0$, and $w$$=$$1$, respectively. Then the free energy, 
$F(\beta)$$\equiv$$-{{1}\over{\beta}} W(\beta)$, 
at one-loop order in this background is given by the expression 
\cite{deconf}:
\begin{equation}
F = \beta^{-1} \int_0^{\infty} {{dt}\over{4t}} 
(2\pi t)^{-12} \eta(it)^{-24} \left 
 ( 1 + 2 \mp {{\eta(it)^{12}}\over{ \Theta_{00}(0;it)^{12}}} \right ) 
\sum_{n \in {\rm Z} , \Delta w = 0, 1} q^{ R^{\prime 2} /\pi^2 \alpha^{\prime} 
+ \alpha^{\prime} \pi^2 n^2 /\beta^2 + (\Delta w)^2 \beta^2/4\pi^4 \alpha^{\prime} }
\quad .
\label{eq:amp2t}
\end{equation}
The classical term in the worldsheet action
can now be removed without appearance of the tachyon. In other words,
$R^{\prime}$ can be taken smoothly to below string scale distances without
a tachyonic instability.
Approached from the high temperature end, the stretched strings turn 
massless at a critical temperature, $T_c$, given by 
$T_c^2$$=$$1/4\pi \alpha^{\prime}$. Keeping $T$$=$$T_c$ fixed, we can
perform a $T$-duality, $R$$=$$\alpha^{\prime}/R^{\prime}$. Working
in the $T$-dual theory with coincident D25branes on a compactified circle, 
we can smoothly examine the distance regime far below the string scale in the 
primed coordinate: $R$$\to$$\infty$, $R^{\prime}$$<<$$\alpha^{\prime 1/2}$.
Conversely, this is the noncompact limit from the viewpoint of the original 
coordinate, $X_{25}$.  The gauge symmetry is enhanced to $SO(2^{13})$ at 
$T_c$; this string state is free of both tachyon, and dilaton tadpoles, 
thereby stable at temperatures of order the string scale and beyond.
To summarize, the critical temperature above which the bosonic 
open and closed string 
state, with as many as $25$, arbitrarily small, compact spatial dimensions, 
is well-defined is $T_c^2$$=$$1/4\pi \alpha^{\prime}$. It is this 
bosonic string state with enhanced gauge symmetry, and fixed critical 
temperature, $T$$=$$T_c$, we will be interested in while matching to a 
bosonic matrix theory analysis. We comment that $T_c$ is also the self-dual
temperature of the closed bosonic string. Our conjecture is 
that the classically stable string state at $T_c$ represents the IR stable 
quantum ground state for the classically unstable, or metastable, bosonic 
string states. As predicted at weak coupling in open and closed string 
perturbation theory. 

\vskip 0.2in
We comment that it may be possible to test this conjecture using string 
field theory methods \cite{stfldth,tachyft}. From the perspective of 
matrix theory, this is not
very significant since our main focus is on the conjectured equivalence 
of the classically stable bosonic string ground state at $T$$=$$T_c$ with the 
planar limit of bosonic matrix theory. This equivalence can be tested by comparison of 
the higher derivative interactions in either case. If this conjecture turns 
out to be true, it would represent a remarkable advance in our understanding 
of the matrix framework for M theory. 

\section{Classical Matrix Actions and Matrix Algebras}

We now move on to a discussion of finite $N$ matrix algebras. We begin with the 
classical matrix action. We will construct it analogous to the well-understood 
example of a Lagrangian in classical field theory, where we begin by picking a 
(spacetime)$\times$(internal) symmetry group. Recall that the fields in the Lagrangian
are required to transform in irreducible multiplets (irreps) 
of the spacetime Lorentz group. In 
addition, they may carry nontrivial charge under one, or more, internal symmetries.
Translational and rotational symmetries, save for Lorentz boosts, could also
be broken. This case implies the
inclusion of certain fields in the Lagrangian with nontrivial momentum in
one or more spatial direction. The procedure by which we arrived at
our proposed action for Matrix Theory in \cite{spont} is simply the 
supersymmetric analog of this analysis, except that we work directly with 
$U(N)$ matrix variables. For clarity, we will only discuss the simpler
bosonic matrix theory in what follows. 

\vskip 0.1in
The fundamental variables in a bosonic matrix action are objects living in 
the $N^2$-dimensional adjoint representation of the unitary group $U(N)$, 
or any of the higher-dimensional irreducible multiplets obtained from the 
decomposition into irreps of the tensor product of an arbitrary number of 
adjoint multiplets. We begin with bosonic variables living in either the 
adjoint, 
antisymmetric traceless, symmetric traceless, or singlet, representations 
of $U(N)$. These irreducible representations appear already in the 
decomposition of the tensor 
product of two adjoints. Notice that although the individual components of 
a bosonic matrix are variables taking value in the field of ordinary real 
(complex) numbers, the matrix itself is a noncommuting object obeying the 
rules of $U(N)$ matrix multiplication.\footnote{The failure to give a 
clear prescription 
for the ordering of $U(N)$ matrix variables at the outset resulted in 
considerable confusion in subsequent matching calculations for graviton 
scattering between the BFSS M(atrix) theory and eleven-dimensional 
supergravity \cite{diner}.} Thus, the ordering of matrices
within a composite product is of crucial importance. 
An unambiguous
prescription for matrix ordering is necessary prior to any meaningful 
analysis of matrix actions.

\subsection{Manifest $SL(26,R)$$\times$$U(N)$ Covariance}

In \cite{spont}, we 
pointed out that an unambiguous prescription for the ordering of matrices in a
composite operator is given by requiring each $U(N)$ multiplet to simultaneously 
transform covariantly under the $SL(26,R)$ subgroup of the finite $N$ Lorentz group. 
Thus, in analogy with spacetime Lagrangians, each term in the classical matrix 
action will be required to be an invariant of the finite $N$ Lorentz group. In 
addition, each matrix variable is required to live in some finite dimensional 
representation of the group $U(N)$$\times$$SL(26,R)$. The construction of invariant 
matrix actions then proceeds by the Noether procedure, familiar from classical 
field theory.

\vskip 0.1in
We work in the first order formalism for Einstein gravity. 
The basic objects are the vierbein, an array of $d^2$ $U(N)$
matrix variables, $ E_{\mu}^a $, subject to the $d$ constraints, 
$E^{\mu a} E_{\mu}^b$$=$$\eta^{ab}$, and the nonabelian vector 
potential, $A_{\mu}$.  The dimensionality, $d$, of the 
auxiliary (flat) tangent
space is undetermined in the classical theory, allowing for
the possibility of ground states with an arbitrary number of
noncompact dimensions $<$$d$. We will assume, however, 
the Minkowskian signature $(-,+,\cdots,+)$ in the
tangent space, coordinatized by $d$ real-valued parameters, $\xi^a$,
and with box-regulated volume $V_{d}$. 
Thus, associated with each point in tangent space is a whole 
$d(d$$-$$1)$$N^2$ unrestricted variables contained in the vierbein,
encapsulating information about
the background spacetime geometry of some large $N$ ground state. 

\vskip 0.2in
The individual $ E_{\mu}^a $ transform covariantly under 
$U(N)$$\times$$SL(26,R)$. Recall that $E_{\mu}^a$, $E_{\mu a}$, 
transform, respectively, as (1,0) and (0,1) bispinors under
$SL(26,R)$; each is in an adjoint multiplet of the $U(N)$. 
These $d(d$$-$$1)$ independent $U(N)$ matrices may be
assembled into composite matrix operators corresponding to 
the physical graviton, antisymmetric twoform, and scalar
multiplets of the finite $N$ Lorentz group.
Thus, the metric tensor, $G_{\mu\nu}$, is a $(1,1)$ tensor 
under $SL(26,R)$, but is expressed as a bilinear composite of
$U(N)$ adjoints, $E_{\mu}^a E_{\nu a} $. 
Likewise, for the physical antisymmetric twoform potential,
$E_{\mu}^a E_{\nu}^{b} \epsilon_{ab}$, and the physical 
dilaton, $E^{\mu a} E_{\mu a} $. 
The nonabelian vector potential also transforms covariantly under
$U(N)$$\times$$SL(26,R)$. We have an $d$$\times$$d_{\rm G}$-dimensional 
array of $U(N)$ adjoint multiplets, $A_{\mu}^i \tau^i $. Each 
element of the array is a Lorentz vector, transforming in the 
adjoint representation of the nonabelian gauge group $\rm G$.
The choice of gauge group is, apriori, arbitrary.
It is however uniquely determined to be $SO(2^{13})$
once we require that the planar spacetime continuum limit 
of the Bosonic Matrix Theory
is an anomaly-free low energy effective field theory. 

\vskip 0.2in
Finally, we can give Lorentz invariant definitions for infinitesimal 
length and volume elements in a given vector space as follows. We define the 
length of a $d$-vector $d V^{\mu}$$\in$${\cal V}$, where each element of
$\cal V$ is an infinitesimal $N$$\times$$N$ matrix, as follows:
\begin{equation}
|dV|^2 ~=~ {\rm Tr}_{U(N)} ~ E^a_{\mu} dV^{\mu} E_{\nu a} dV^{\nu}  \quad ,
\label{eq:length}
\end{equation}
where the trace is over $U(N)$ indices. Similar definitions can be given 
for the $p$-th volume form, $p$$=$$2$, $\cdots$, $26$, in any 
finite-dimensional vector space. The result in each case is an 
ordinary real number. It is evident that this definition corresponds 
to the usual definition of length and volume invariants of the $d$-dimensional 
Diffeomorphism group in the large $N$ limit. 

\vskip 0.2in
Notice that, thus far, all of the required matrix variables in the
bosonic matrix action have been expressed in terms of $U(N)$
adjoints, for e.g., $A_{\mu}$, or appropriate 
composites of $U(N)$ adjoints, namely, the $E^a_{\mu}$. It is often
helpful to express the classical action in terms of the composite
matrix variables 
since these correspond directly to the fields appearing in the 
low energy spacetime Lagrangian: the scalar dilaton, $\Phi$, 
symmetric two-form or metric, $G_2$, and antisymmetric two-form, $A_2$. 
When the large $N$ ground state carries nontrivial Dbrane charges,
additional higher dimensional $U(N)$ irreps will be required in the matrix 
action. These new matrix variables cannot be expressed as composites of 
adjoints. Recall that, due to the presence of nonabelian gauge fields, 
the low energy spacetime Lagrangian of the bosonic string contains mixed 
Chern-Simons terms coupling $A_1$ to antisymmetric $p$-form potentials, 
$C_{[p]}$, with $p$ $=$ $0$, $\cdots$, $25$. Such Chern-Simons terms have 
their counterparts in the matrix theory action. Given an understanding 
of the manifest symmetries desired in the classical matrix action, let 
us proceed with implementing the Noether procedure and with examining the 
detailed form of the action. 

\subsection{Extended Yang-Mills Symmetry}

\vskip 0.1in
We begin with the definition of the gauge covariant derivative.
At finite $N$, the gauge covariant derivative operator takes the 
matrix form, $ D_a $$=$$ \Omega_a $$+$$  g A_a^i \tau^i $. The symbol 
\lq\lq $;$" will be used to denote left-multiplication by $\Omega_a$,
It acts as the general covariant derivative of the object to its immediate
right in the planar limit: $\Omega_a $$\to$$\partial_a $, in a
flat spacetime background. Recall that $\Omega_a$ transforms as
a Lorentz vector, in the $(\half, \half )$ representation of SL(26,R). 
The Riemann curvature scalar is expressed in 
the form:
\begin{equation}
{\cal R} [E] = (\Omega_b E^{b\lambda}) (\Omega_a E^a_{\lambda } ) 
-  (\Omega_a E^{b\lambda}) (\Omega_b E^a_{\lambda})   
+  E^{a\lambda} (\Omega_a E^b_{\sigma }) (\Omega_b E^c_{\lambda})   E^{\sigma}_c 
-  E^{a\lambda}  (\Omega_a E^c_{\lambda }) E_{c}^{\sigma} (\Omega_b E^b_{\sigma})  
\quad .
\label{eq:mcurvy}
\end{equation}
$\cal R$ is quadratic in covariant derivatives of the $E_a^{\mu}$.
Notice that operator ordering has been determined by requiring that 
each term in the matrix action is invariant under the finite $N$ 
matrix Lorentz transformations \cite{spont}:
\begin{eqnarray}
\delta_L (U_a \Phi ) =&&  [L_a^c , U_c ] \Phi 
, \quad  
\delta_L ( \Phi U^a) = - \Phi [U^c , L_c^a ]  
\cr 
\delta_L (U_a V_b ) =&&  [L_a^c , U_c ] V_b +
  U_a [L^c_b , V_c] \cr 
\delta_L (U^a V_b ) =&&  
  U^a [L^c_b , V_c] - [U^c , L^a_c ] V_b 
\quad ,
\label{eq:lor}
\end{eqnarray}
for arbitrary Lorentz vectors $U$, $V$, and scalar $\Phi$.
The parameter of Lorentz transformations, $L_{ab}$, is an array of infinitesimal 
$N$$\times$$N$ matrices, antisymmetric under the interchange of indices $a$,$b$.

\vskip 0.15in
The Yang-Mills and antisymmetric threeform field strength can be written in matrix
form as follows: 
\begin{eqnarray}
F^i_{ab} =&& \Omega_a A^i_{b} - \Omega_b A^i_{a} +  gf^{ijk} A^j_{a} A^k_{b}
\cr
H_{abc} =&& 
\Omega_{[a} A_{bc]} - X_{abc} \equiv
\Omega_{[a} A_{bc]} -
 2^{1/2} ~ {\rm tr}_{ijk} ~ (\delta_{ij} A^i_{[a}\Omega_{b} A^j_{c]}
  - {{2}\over{3}} A^i_{[a} A^j_{b} A^k_{c]} )
\quad .
\label{eq:m3form}
\end{eqnarray}
With this definition, the kinetic terms for both $F$ and $H$ take the standard form 
while incorporating the extended Yang-Mills symmetry. The mixed Chern-Simons 
terms are responsible for the coupling between matrix gauge potentials of 
different rank. Each of the matrix potentials transforms in an irrep of $U(N)$,
respectively, adjoint, and antisymmetric tensor.
This argument extends to all of the higher rank matrix potentials. 
The appropriate $U(N)$ irrep required is identified by 
consulting a good group theory table, requiring that the kinetic 
term for the $p$-form potential in the matrix action is manifestly 
invariant under $U(N)$$\times$$SL(26,R)$. In each case, we define
the shifted field strength in order that the mixed Chern-Simons terms are 
incorporated naturally (see the discussion in Chap.\ 12 of \cite{polbook}). 

\vskip 0.15in
For example, consider a ground state of bosonic matrix theory which couples 
to an even rank $p$-form matrix gauge potential, $C_{[p]}$.  Due to nontrivial 
$C_{[p]}$ charge, the bosonic matrix theory action contains a term from
the series given below. Note that the $C_{[p]}$ live in independent, 
increasingly higher rank, irreps of $U(N)$:
\begin{eqnarray}
 S_0 =&& ({\cal F}_1 )^2 , \quad {\cal F}_{a} = \Omega_{a} C_{0} - A_{1} 
\cr
S_{2} =&& \half {\cal F}_3 \wedge {\cal F}_3 ,
\quad {\cal F}_{abc} = \Omega_{[a} C_{bc]} - X_{abc}
\cr 
 && X_{abc} = 2^{1/2} {\rm tr}_{ijk} ~ (\delta_{ij} A^i_{[a}\Omega_{b} A^j_{c]}
  - {{2}\over{3}} f_{ijk} A^i_{[a} A^j_{b} A^k_{c]} )
\cr
S_4 =&& \half {\cal F}_5 \wedge {\cal F}_5 ,
\quad\quad {\cal F}_{abcde} = \Omega_{[a} C_{bcde]} - X_{abcde}i ,
\cr 
 && X_{abcde} =  {\rm tr}_{ijk} ~ (\delta_{ij} A^i_{[a}\Omega_{b} A^j_{c} C_{de]}
  - f_{ijk} A^i_{[a} A^j_{b} A^k_{c} C_{de]} ) + \cdots  
\cr
\cdots && \cdots
\cr  
S_{26} =&& \half {\cal F}_{27} \wedge {\cal F}_{27} ,
\quad\quad \cdots \quad .
\label{eq:stre}
\end{eqnarray}
We should emphasize that that the necessity for independent $U(N)$ irreps for 
each of the matrix potentials in the action for Matrix Theory is forced upon 
us by the properties of the symmetry algebra. This is not unlike the case of
open and closed string theory; while open strings do produce closed strings 
at the loop level, the renormalization of open and closed string coupling are 
{\em independent} results (see the detailed explanation in 
\cite{poltorus,ncom}). Thus, at the quantum level, the open and closed string 
sectors contribute independent degrees of freedom in string theory.  
In particular, it is not true that \lq\lq gauge theory contains gravity" in 
String Theory, an unfortunate misconception that appears in some of the recent
literature. 

\vskip 0.2in
The existence of Extended Yang-Mills symmetries in String/M theory simply implies 
the necessity for still further independent degrees of freedom associated with 
each of the higher rank gauge potentials. These are the Dbranes and, possibly,
brane-antibranes, Mbranes, and other solitons. In Matrix Theory, we will find 
that all of the higher rank matrix potentials associated with Dbranes are 
represented democratically in the matrix effective action.

\subsection{Quantum Effective Action and Dbrane Democracy}

\vskip 0.2in
We will now make an important observation.
Notice that the commutator of the first member of the series given
in Eq.\ (\ref{eq:stre}) with the generator of 
matrix Lorentz transformations, $[C_o,L_{ab}]$, 
is nontrivial.
This implies that finite $N$ Lorentz invariance
requires the ground state to couple to a two-form potential, 
$C_{[2]}$, also an $U(N)$ matrix. The argument can be iterated to conclude that
coupling of the ground state to any one even rank potential implies, 
as a consequence of Lorentz invariance, the coupling to {\em all} even rank 
matrix potentials, with $p$$<$$25$. Conversely, if we begin with the ground 
state with nontrivial coupling to a $26$ form matrix gauge potential, 
$C_{[26]}$, the commutator with $L_{ab}$ results in nontrivial 
coupling to the full spectrum of even $p$-form potentials, with 
$p$$<$$25$. 

\vskip 0.2in
In addition, since we include a one-form Yang-Mills potential in 
the matrix action, the
nontrivial $U(N)$ matrix commutator, $[A_a , C_{26}]$, has an
expansion in terms of couplings to the {\em odd} rank $p$-form
potentials, where $p$ takes all odd integer values less than
$26$. Thus, the requirement of invariance under the gauge symmetry
mediated by the oneform vector potential in addition to Lorentz
invariance, implies a coupling to all of the remaining {\em odd} 
rank potentials.  We emphasize that, as regards the finite $N$ algebraic 
structure, matrix Dbrane democracy follows even when the one-form
vector field, $A_a$, is {\em abelian}. However, in the presence of 
a Yang-Mills symmetry, the finite $N$ matrix algebra has a beautiful
spacetime remnant, in the form of mixed Chern-Simon terms that survive 
in the planar continuum action. We have recovered a well-known result 
from the spacetime low energy effective field theory analyses of 
Dbranes. Namely, that Dpbrane charge conservation must be carefully 
defined so as to correct for the effect of mixed Chern-Simons terms
in the effective field theory \cite{douglas}. The origin of this 
mixing lies in closure under the finite $N$ Lorentz and extended 
gauge symmetry algebras. 

\vskip 0.2in
We move on to a discussion of the quantum matrix effective action,
exhibiting manifest invariance under the full 
$SL(26,R)$$\times$$U(N)$$\times{\cal G}$.
We have the following dimensionless free parameters available to us in the 
unoriented bosonic string theory: $g_0$, or the Yang-Mills coupling, and the 
constant parts of the background fields, $A_{ab}$, $F_{ab}$, and $\Phi$, assuming 
an expansion about flat Lorentzian spacetime. Notice that the gravitational, or 
closed string, coupling has been traded for the background vev of the dilaton scalar. 
In addition, we have the antisymmetric higher rank $p$-form matrix potentials 
described above.

\vskip 0.1in
We work in the matrix analog of the modified axial gauge for quantization 
of the one-form potential, preserving the residual gauge invariances of axial
gauge and other symmetries of the effective action.
This corresponds to a $1/N$ expansion about the timelike Wilson line 
background. Here, $\bf 1$ denotes the identity $N$$\times$$N$ matrix 
in $U(N)$ space, and the $t^A$, $A$$=$$1$, $\cdots$, $2^{13}$, are the 
hermitian generators of the Yang-Mills group $SO(2^{13})$. Thus, the 
background potentials are mutually-commuting matrix variables. We consider
expanding about the following classical background of bosonic
matrix theory: 
\begin{equation}
{\bar{E}}^a_{\mu} {\bar {E}}_{\nu a} = (\eta_{\mu\nu} + \kappa h_{\mu\nu} ) {\bf 1} ,
\quad 
{\bar{A}}_{0}^A t_A = \beta^{-1} {\bf 1} , \quad 
{\bar{A}}_{i}^A t_A = 0 , ~ i = 1,\cdots , 25 \quad .
\quad 
{\bar{C}}_{[p] } =  c \delta_{\mu 25} {\bf 1}  \quad .
\label{eq:25braneE} 
\end{equation}
As explained earlier, the quantum effective action will contain induced charges 
for the full spectrum of $p$-form matrix potentials. Moreover, the charges and
relative normalizations of all terms in the matrix effective action are 
entirely determined by symmetry given the classical input from above.
This large $N$ ground state describes a flat 26-dimensional space with compact 
(Euclidean) time, and a timelike Wilson line. There are $2^{12}$ space-filling 
D25branes and, consequently, 
a constant 26-form antisymmetric tensor potential. The propagating degrees of 
freedom are 26-dimensional nonabelian gauge fields and gravitons, originating
in the linearized matrix-valued perturbations: ${\tilde{A}}_a$, 
${\tilde{h}}_{ab}$, ${\tilde{\phi}}$, ${\tilde{A}}_{ab}$, 
and ${\tilde{C}}_{[p]}$. 

\vskip 0.1in 
\noindent The quantum
path integral for bosonic matrix theory is given by an
expression of the form:
\begin{equation}
{\cal Z} = \int {{[dE][dA][dC]}\over{{\rm Vol}(SL(26,R) \times U(N) \times {\rm G} ) }}
 e^{- {\cal S} (E,A,C; \{ g , {\bar{A}} \} )} \equiv 
 \exp \left (- {\cal S}_{\rm eff}[{\tilde{E}},{\tilde{A}},{\tilde{C}}] \right )
\quad , 
\label{eq:path}
\end{equation}
where $\{ g; {\bar{A}} \}$ denotes the set of free couplings and parameters describing the 
large $N$ background about which we expand using linearized perturbation theory. 
But as we will see in a moment, the matrix
quantum effective action is fully determined by symmetry alone. Thus, it 
is possible to write down the result for ${\cal S}_{\rm eff}$ directly:
\begin{equation}
{\cal S}_{\rm eff} =  
e^{\Phi} ( \half g^2 {F}^{ab} F_{ab} + 
\half \kap e^{ \Phi} \sum_{p=1}^{27} p ~ {\cal F}_{[p]} \wedge {\cal F}_{[p]}  ) 
  + \half \kap ( {\cal R}
              - 4 {\Omega}^{a} \Phi  \Omega_{a} \Phi + 
3 e^{2 \Phi}  {H}^{abc} {H}_{abc} ) \quad . 
\label{eq:bmat}
\end{equation}
We emphasize that there is no choice involved in either including,
or excluding, the higher rank potentials: in their absence, the ground 
state would have no gauge fields, and the theory has no IR stable ground
state. Thus, we must incorporate Yang-Mills fields. Due to finite $N$
Lorentz invariance, this automatically implies inclusion of all of the 
terms appearing in Eq.\ (\ref{eq:bmat}).  

\vskip 0.1in
Notice also that there is no ambiguity in the relative normalizations of the
kinetic terms of the higher rank $p$-form potentials, as a consequence of  
the finite $N$ Lorentz-extended-Yang-Mills invariance. Thus, the remnant
mixed Chern-Simons terms obtained in the planar limit are unambiguously
determined by the extended $p$-form-Yang-Mills gauge symmetry. This is 
a clear prediction for the quantum string effective action to which it is
matched. Notice that, unlike the unoriented bosonic string effective action, 
prior to taking the large $N$ limit, all of the terms required by the
finite $N$ symmetries are present in the matrix action: there is both an
$H_{[3]}$, and an ${\cal F}_{[3]}$. This fact will be noteworthy in the 
supersymmetric case \cite{spont}. Notice that the matrix potential $A_{ab}$ 
arises in the vierbein (gravity) multiplet; $C_{ab}$ occurs as a 
consequence of 
Yang-Mills invariance. Owing to the fact that we begin with a ground state 
charged under $C_{[25]}$, and invariant under the gauge symmetry 
mediated by the one-form potential $A_a$. 

\vskip 0.1in
In the fully democratic ground state with $SO(2^{13})$ Yang-Mills 
symmetry, it is possible to verify invariance of the quantum effective
action, ${\cal S}_{\rm eff}$, under the 
finite $N$ matrix algebra,
namely, under $U(N)$$\times$$SL(26,R)$$\times$$\cal G$. 
For completeness, we write these down explicitly as in \cite{spont}.
We will introduce an infinitesimal
hermitian matrix, ${\rm L}_{ab}$,
antisymmetric under the interchange of tangent space indices
$a$,$b$. Keeping terms up to linear in ${\rm L}_{ab}$, it is easy
to verify that each term in ${\cal S}$ is invariant under
the matrix transformations:
\begin{eqnarray}
\delta A_a  =&& [ {\rm L}_{a}^c , A_c ]
\cr
\delta A_{ab}  =&& [ {\rm L}_{a}^e , A_{eb} ] + [ A_{ae}, {\rm L}^e_b ]
\cr
\cdots = & \cdots
\cr
\quad \delta ( \Omega_{a} \Phi ) =&& [ {\rm L}_a^c , \Omega_c ] \Phi
- \Omega_c {\rm L}^c_a \Phi ,
\cr
\delta ( \Phi \Omega^a ) =&& - \Phi [ \Omega^c , {\rm L}_c^a ] +
 \Phi {\rm L}^a_c \Omega^c
\label{eq:lorent}
\end{eqnarray}
Likewise, consider a $d_G$-plet of infinitesimal real matrices,
$\{ \alpha^j \}$, each of which takes diagonal $N$$\times$$N$ form.
Here, $d_G$ is the dimension of the nonabelian gauge group with
hermitian generators $\{ \tau_j \}$. We can verify that every term in
$\cal S$ is invariant under the Yang-Mills transformations:
\begin{eqnarray}
\delta (g A_a^j \tau^j ) =&& [ \Omega_a , \tau^j \alpha^j ]
\cr
\delta ( \Omega_{a} \Phi ) =&& i \tau^j \alpha^j \Omega_a \Phi
, \quad \quad \delta ( \Phi {\hat{\Omega}}_{a} ) = - i \tau^j \Phi {\hat{\Omega}}_a
 \alpha^j \quad .
\label{eq:gaugena}
\end{eqnarray}
Finally, it is easy to verify invariance under the higher rank 
symmetries, since the kinetic term for the field strength has 
been written in standard form by defining an appropriate shift.

\section{Conclusions}

\vskip 0.1in
Bosonic Matrix Theory has one dimensionless free parameter, $g$, and one intrinsic
mass scale, $\alpha^{\prime}$. We can, of course, trade these for the D25brane
tension and Yang-Mills gauge coupling.\footnote{In a realistic ground state 
of the supersymmetric Matrix theory, these parameters are determined by 
matching with the physical
coupling unification scale, and the strength of the unified Yang-Mills-gravity 
coupling.} In addition, we have available a 
broad choice of dimensionless, and dimensionful, scalings of the spectrum
of background fields, namely, the spacetime metric,
Yang-Mills and antisymmetric two-form potential, and higher rank 
$p$-form matrix potentials.
As was pointed out by us in \cite{spont},
such dimensionful, multiple-scaled, large $N$ limits of Matrix Theory
are in one-to-one
correspondence with the modified low energy effective field theory limits 
distinct from the original zero slope limit of string theory \cite{malda}. 
Such gauge-gravity effective duals arise precisely as a consequence of 
holding one, or more, additional mass scale fixed while taking the 
zero slope limit of a string theory. In matrix theory language, such a 
modification of the large $N$ limit is simply an extension 
of the notion of the $(g,N)$ double-scaling limit familiar from the 
one-matrix model \cite{doubles}. 

\vskip 0.1in
It should be noted that the $1/N$ corrections to the planar limit of the Matrix 
Theory quantum effective action contain higher derivative terms which may be 
compared with those obtained in the leading orders of the $\alpha^{\prime}$ 
expansion of the string effective action. This should provide new insight into
quantum corrections to the classical string theory predictions for the 
widely-studied gauge-gravity dualities of M Theory \cite{malda}. In connection
to this subject, we should also mention the role of string loop corrections
to the quantum effective action. In this paper, we have naturally focussed 
our attention on results which are meaningful at weak, but non-vanishing, 
string coupling, and in an IR stable ground state. In the fully supersymmetric
case, studied by us earlier in \cite{spont}, there is in
addition, the possibility of incorporating weak-strong coupling duality. 
Specifically, we can match with the weak-strong dual heterotic-type I string 
theory limits but, also, the self-dual type IIB string theory limit. The matrix 
theory analog of the IIB string theory should be of great interest in the context 
of matrix Dbrane democracy. Direct comparison with some of the results of the 
Ishibashi-Kawai-Kitazawa-Tsuchiya matrix model \cite{ikkt} may be possible. 
It would be particularly interesting to develop analogous techniques to study 
genuinely nonperturbative phenomena, such as the matrix computation of Wilson Loop 
correlators or the dynamical selection of the quantum ground state \cite{ikkt}, 
in our full-fledged proposal for Matrix Theory \cite{spont}.

\vskip 0.1in
We emphasize that it is very important to study the full scope of the correspondence
we have described above, since upon its validity hinges the success of our proposal
for Matrix Theory \cite{spont}: does our matrix framework account for all of 
the known facts about weak-strong-dual effective field theory limits of M 
theory? Are there any known backgrounds which cannot be incorporated within the 
Matrix Theory framework? We remind that reader that it was an inability to
incorporate certain backgrounds with fewer than five noncompact dimensions 
\cite{bfss,diner}--- at least in testable form, that led to the eventual demise of 
the BFSS M(atrix) theory. Although we have emphasized the action formulation 
at the outset in our work, because of its pedagogical value, it is obvious 
that there is nothing in our algebraic framework that {\em requires} it. In 
formulating the arbitrary matrix Dbrane state in section 3.2, we have described 
it in terms of a closed algebra. Likewise, although we are biased in this paper 
towards backgrounds of Matrix Theory whose planar limits contain propagating 
Yang-Mills and gravitational fields, neither is {\em necessary} in the arbitrary 
Dbrane state. Thus, little string theories \cite{dvv} are expected to be
compatible with our framework. It is well-known that type I states with additional 
Dbrane-antiDbrane pairs likewise have a simple description in terms of a closed 
algebra \cite{kth}. As is also true of states with NS fivebranes, and  
more complicated M-brane solitons. In summary, we see no insurmountable 
difficulties in accommodating the most general background of M theory in our
proposed algebraic framework.

\vskip 0.1in
It remains a matter of great interest to develop the Hamiltonian formulation of 
Matrix Theory, analogous to the Banks-Fischler-Shenker-Susskind M(atrix) Theory
framework \cite{bfss}. In \cite{spont}, we have emphasized that the notion that
the \lq\lq fundamental" degrees of freedom in Nature correspond to zero-dimensional 
matrix variables is a fact best appreciated from the representation-independent 
viewpoint of Dirac's Matrix Quantum Mechanics. From this perspective, it is 
a perfectly natural assumption. Since we have a comparatively clear understanding 
of propagating degrees of freedom that fill both space and time, namely, quantum 
field theory, we can simplify our considerations, eliminating space by invoking
translational invariance to \lq\lq reduce" the dynamical degrees of freedom to
those living on a single spatial point \cite{ek}. 
This leads to a Hamiltonian framework with a uniquely 
specified time. Or, we can eliminate both space and time, leading naturally
to the pre-geometrical 
notions of noncommutative geometry \cite{ek}. 
Hence the close connection between
noncommutative geometry and the reduced matrix models, exploited in the work of
\cite{ek,ikkt}. 

\vskip 0.1in
This is the main idea underlying Eguchi-Kawai reduction \cite{ek}. It is the framework 
within which we have cast our formalism for 
Matrix Theory, and the Action Principle remains an important tool 
for its study. We emphasize that semi-classical investigations which will undoubtedly 
shed much light on this framework, can make full use of the related methodology 
of Hamilton-Jacobi theory. We should comment here that it would be very 
interesting to examine the matrix analog of the classical Dirac-Born-Infeld action 
for Dbranes. In particular, this could lead to a much simplified discussion of the 
vexing nonabalian DBI action describing the limit of coincident Dbranes \cite{tseyt}. 

\vskip 0.1in
Although we have emphasized the arbitrary Dbrane state above, we find it encouraging 
that the fully democratic quantum ground state of Bosonic Matrix Theory with a finite,
and anomaly-free, planar limit corresponding to the zero slope limit of the 
unoriented open and closed bosonic string, has a simple and elegant description 
within the action formulation. Our investigations have their obvious parallel in
an analogous state of the nonperturbative type I-I$^{\prime}$-heterotic theory and 
its matrix analog \cite{spont}. Future analyses which explore the 
full nonperturbative dynamics of Matrix Theory will tell us whether these 
observations are prescient. 

\bigskip
\bigskip
\bigskip

\noindent{\bf Acknowledgment:} This research was funded in part by grant 
NSF-PHY-9722394 from the National Science Foundation under the auspices of 
the CAREER program.

\vskip 0.3in 
\noindent{\bf NOTE ADDED (JULY 2005):} I have corrected one inexplicable (silly) typo 
in the draft;
$SL(2,C)$ in place of, the obvious, 
$SL(26,R)$. I wish somebody had brought this to my notice. I consider this a beautiful
paper, with many ideas, whose impact has still to be realized in the full 
supersymmetric matrix proposal for nonperturbative String/M theory. The expression for the
\lq\lq free energy"
in the discussion in section 2 is obviously only part of the answer, missing the unoriented 
and closed string contributions, clarified already above Eq.\ (3).  The
\lq\lq inescapable" conclusion at the top of page 4 refers only to the type II closed 
superstrings in flat spacetime with trivial Ramond-Ramond sector. \lq\lq Extended"
Yang-Mills symmetries just means higher rank pform gauge potentials.

\bigskip
\medskip
\medskip

\noindent{\bf Appendix: Low Energy Spacetime Action of the Unoriented Bosonic String}

\vskip 0.1in
For completeness, we list our conventions and recall the precise form of the 
Lorentz invariant spacetime Lagrangian describing the low energy limit of the 
unoriented open and closed bosonic string theory \cite{polbook}. We use the 
first order vierbein formalism for Einstein gravity, and work in
the weak field approximation, perturbing about the fixed background metric, 
$G_{\mu\nu}(x)$$=$$\eta_{\mu\nu}$$+$$2 \kappa h_{\mu\nu}$, as in \cite{br}. 
The dimensionful 
coupling $\kappa$ is the 26-dimensional Newton's constant, $(8\pi G_{\rm N})^{1/2}$.
For full generality, we could
allow for a possible constant background electric field
pointing in the spatial direction $X^{24}$, as well as a constant background $B$ 
field. Thus, the two-dimensional space $(X^{23},X^{24})$ could be noncommutative. 
This range of backgrounds will permit us to make appropriate 
comparisons with a broad range of results from perturbative open string theory. 
\vskip 0.1in
The physical fields of the gravity multiplet consist of the graviton, dilaton 
scalar, and antisymmetric twoform tensor field. The vierbein is denoted 
$e^a_{\mu}(x)$, where $a$$=$$0$, $\cdots$, $26$, parameterizes the flat 
local tangent space, and $\mu$ is the spacetime index. The physical composite
fields can be expressed as follows:
\begin{equation}
g_{\mu\nu}(x)  = e^a_{\mu}(x) e^b_{\nu}(x) \eta_{ab}, \quad
\phi(x)  = e^a_{\mu}(x) e^{\mu}_a (x) , \quad
b_{\mu\nu}(x)  = e^a_{\mu}(x) e^b_{\nu}(x) \epsilon_{ab}
\quad .
\label{eq:metric}
\end{equation}
The dilaton vev and gravitational coupling appear in the action only when multiplied
together, $\kappa e^{-{\bar{\phi}}}$. Thus, any finite rescaling, or renormalization, 
of the closed string coupling should be understood as a change in the vacuum expectation
value for the dilaton field \cite{polbook}. The Yang-Mills multiplet is composed of 
the vector potential,
$f^{ijk}A^{i}_{\mu}(x)$, $i$$=$$1$, $\cdots$, $d_G$, transforming in the
adjoint representation of the orthogonal group $SO(2^{13})$. The choice of gauge 
group can be determined by requiring the absence of all worldsheet anomalies or, 
equivalently, all gauge and gravitational spacetime anomalies \cite{polbook}.
The Yang-Mills coupling is dimensionless and is to be identified with the open 
string coupling, $g$ $=$ $g_{\rm open}$ \cite{polbook}. We emphasize that the 
perturbative renormalizations of open and closed string couplings are known to 
have independent origin \cite{polbook,ncom}. This result implies the analogous 
independence of the perturbative renormalizations of the gauge and gravitational 
couplings appearing in the low energy spacetime effective Lagrangian.

\vskip 0.1in
In the weak field approximation, the Ricci curvature tensor takes the simple form:
\begin{equation}
R_{\mu\nu} ~=~ 
\kappa \left ( \Delta h_{\mu\nu} - \partial^{\lambda} \partial_{\nu} h_{\lambda \mu}
+ \partial_{\mu} \partial_{\nu} h^{\lambda}_{\lambda} \right ) \quad .
\label{eq:ricci}
\end{equation}
where the symbol \lq\lq$;$" denotes the general covariant derivative, allowing for
a nontrivial Christoffel connection. The nonabelian gauge covariant field strength 
takes the familiar form:
\begin{equation}
F^i_{\mu\nu} [A(x)] ~=~ \partial_{\mu} A^i_{\nu} (x) 
- \partial_{\nu} A^i_{\mu} (x) +  gf^{ijk} A^j_{\mu}(x) A^k_{\nu} (x)
\quad . 
\label{eq:gauge}
\end{equation}
More generally, the scalar curvature, $R[e(x)]$, can be expressed as:
\begin{equation}
e^{\mu}_a (x) e^{\nu}_b (x) \left [  
e^{b\lambda}_{ ; \nu} (x) e^a_{\lambda ; \mu} (x) 
-  e^{b\lambda}_{ ; \mu} (x) e^a_{\lambda ; \nu} (x)  
+  e^{a\lambda} (x) e^c_{\lambda ; \nu} (x) e_{c}^{\sigma} (x) e^b_{\sigma ; \mu} (x)  
-  e^{a\lambda} (x) e^c_{\lambda ; \mu} (x) e_{c}^{\sigma} (x) e^b_{\sigma ; \nu} (x)  
\right ] .
\label{eq:curvy}
\end{equation}
Thus, allowing up to two time derivatives, the manifestly local Lorentz invariant 
and diffeomorphism invariant spacetime Lagrangian coupled to both a Yang-Mills and 
possible higher $p$-form antisymmetric gauge potential takes the form:
\begin{equation}
{\cal S}_{\rm B} =  - \int d^{d} x e \phi^{d-7} 
\left [ \quart
  F^i_{\mu\nu} [A(x)] F^{i \mu\nu} [A(x)] 
+ {{1}\over{2\kappa^2}} ~ R[e(x)] 
+ {{3}\over{2}} H_{\mu\nu\rho} [e(x)] H^{\mu\nu\rho} [e(x)] 
- \phi \Delta \phi [e(x)] 
\right ] .
\label{eq:boso}
\end{equation} 

\vskip 0.1in
Due to the presence of $\kappa$ dependent terms, the Yang-Mills symmetry is 
automatically extended to include transformations involving the two-form tensor 
field accompanied by ordinary Yang-Mills gauge transformations \cite{br,polbook}.
Furthermore, upon inclusion of a nontrivial $C_{[26]}$ antisymmetric tensor 
potential, the presence of mixed Chern-Simons couplings induced by requiring
closure of the extended gauge symmetry group implies induced $p$-form potentials
for {\em all} of the antisymmetric gauge potentials with $p$$<$$26$. The kinetic 
terms for the gauge potentials can be restored to the canonical form by 
appropriate field redefinitions in the antisymmetric $(p$$+$$1$)-form field 
strength tensors. For the threeform field strength, the shift takes the form
\cite{br}:
\begin{equation}
H_{\mu\nu\rho} \equiv 
\partial_{[\mu} A_{\nu\rho]} - X_{\mu\nu\rho} \equiv 
\partial_{[\mu} A_{\nu\rho]} - 
 2^{1/2} ~ {\rm tr} ~ (A_{[\mu}\partial_{\nu} A_{\rho]}
  - {{2}\over{3}} A_{[\mu} A_{\nu} A_{\rho ]} ) 
\quad .
\label{eq:3form}
\end{equation}
We emphasize that in the {\em absence} of the gauge and gravitational couplings--- 
or in the free field limit, the different possible $(p$$-$$1$)-form charges should be 
understood as being independently conserved charges on the vacuum.

\end{document}